\documentstyle[12pt,psfig]{article}
\begin{document}
\sloppy
\thispagestyle{empty}
\begin{center}

{\Large{\bf
Experimental Investigation of New Method of Energy Generation in
Plasma Devices caused by Existence of Physical Space Global Anisotropy.
}}
\vskip20pt

Yu.A.Baurov\footnote{baurov@www.com}, G.A.Beda, \fbox{I.P.Danilenko}, V.P.Ignatko.
\vskip5pt
{\it Central Research Institute of Machine Building,

141070, Korolyov, Pionerskaya, 4, Moscow Region, Russia.
}
\vskip25pt
 ABSTRACT
\end{center}
{\footnotesize
An experimental investigation of a new interaction connected with the
existence of the cosmological vectorial potential, has been carried out.
On its basis, a new method of energy generation with the use of a plasma
generator, has been studied. The experimental results are presented.
}
\pagebreak
\vskip10pt
{\bf 1. Introduction}
\vskip5pt
In Refs. [1-8], a new interaction of the objects in nature,
different from four known interactions (the strong, weak,
electromagnetic, and gravitational ones), has been predicted.
The new force is caused by the existence of the cosmological
vectorial potential ${\bf A_g}$, a new fundamental vectorial constant
appearing in the definition of byuons, new discrete objects
(physical dimension of byuons are as of the electrical charge,
magnetic flux, Dirac's monopole - the same in CGSE system).
According to the hypothesis suggested in Refs. [1,7,8], in
minimization of the potential energy of interaction between the
byuons in the one-dimensional space formed by them, the observable
physical space as well as the world of elementary particles together
with their properties, are arising. In this
model, the masses of the particles are proportional to the
modulus of the summary potential ${\bf A_\Sigma}$, consisting of ${\bf A_g}$ and  the
vectorial potentials of various magnetic sources as of natural
origin (the Earth's and Sun's potentials, etc.) so artificial
ones (for example, the vectorial potentials A of magnetic fields
of solenoids, plasma generators, etc.). The magnitude $|\bf A_\Sigma|$ is
always lesser then $|{\bf A_g}| \approx 1.95\times 10^{11} Gs\cdot cm$ [1-8].

In distinction to the gauge theory (for example, the classic and quantum
field theory), in the model of Refs. [1,7,8] the values of potentials
acquire a physical meaning in tune with the known Aharonov's-Bohm's
effect [9,10] being a special case of the space quantum properties described
in Ref. [8].

The magnitude of the new force is $\displaystyle F\sim N\Delta A\frac{\partial\Delta A}{\partial x}$,
where $\Delta A$ is the difference in changes of $A_\Sigma$ between the
points at which the sensor and test body are placed [1,7,8], $x$ is the
coordinate in space, N is the number of stable massive elementary particles
(electrons, protons, neutrons) in the region of changing ${\bf A_\Sigma}$.
According to the ground-based experiments with high-current
magnets [2-5], with a gravimeter and a magnet attached to it
[11,12], the experiments on investigating the changes in $\beta$-decay rate of
radioactive elements [13,14], and the astrophysical observations [15-17],
${\bf A_g}$ has the following estimated coordinates in the second equatorial
system: the right ascension $\alpha \approx 270^\circ$, the declination
$\delta \approx 34^\circ$.

The new force ejects any substance from the region of diminished $|{\bf A_\Sigma}|$ mainly in
direction of ${\bf A_g}$. The most effective angle between the vector A of a
current system and the vector ${\bf A_g}$ is $150\div140^\circ$ [8].

In the investigations with high-current magnets (magnetic flux density up to 15T), the magnitude
of the new force was equal to $\sim (0.01\div0.08)g$ at the $30g$ mass of the test
body. It was shown in experiments with rotating magnetic discs and an
engine-generator [18-20] that the magnitude of the force $F$ can be
considerably increased when to phase the motion of the body with the
process of formation of the physical space from the byuons (i.e. the
working body is bound to change $A_\Sigma$ by its potential $A$ and move in the
direction of the vector ${\bf A_g}$. Therewith its particles are to rotate in the
proper side). In this case the energy will be extracted from the physical
space through the elementary particles of the working body. The low of
energy conservation in the system "working body-physical space" will be
obeyed. It is known that the main part ($\sim 98\%$) of energy in the Universe
is determined by the "dark" (virtual) matter [21]. The model of Refs.
[1,7,8] describes the phenomenon of the "dark matter" quite satisfactorily.
\vskip10pt
{\bf 2. The experimental installation and technique.}
\vskip5pt
To test the above said, a special stationary plasma generator
with linear discharge (see Fig.1)
has been manufactured at the Central Research Institute of
Machinery. The plasma generator (1) (power $\sim 60kW$, current
$I\sim300A$, voltage $U\sim220V$) is arranged on a rotatable base and can
be turned together with the whole instrumentation through 320
degrees $\varphi$ around the vertical axis (2) and through $90^\circ$ around
horizontal axis. The plasma generator is water-cooled (3). As a
working medium air (4) admixed with argon ($\sim1\%$) was used. A
measuring tube from copper (6) of internal diameter $0.8cm$ was
fastened to the plasma generator by means of a holder (5) at $8cm$
from the nozzle exit section.  The temperature of water drawing
through the tube was $\sim16^\circ C$. The arrangement of the measuring
tube relative to the plume (7) of the plasma generator is shown
in Fig.1.  In the center of the section of the measuring tube at
the inlet of water and in the region of the jet, the junctions of Chromel-Alumel
thermocouples (8) $0.2mm$ in diameter were mounted.
The thermocouples used in the experiments were industrially manufactured
and corresponded to the standard GOST 3044-61 \cite{23}.
The calibration of thermocouples was tested when immersing them into boiling
distilled water ($100^\circ C$).
The percent change in temperature of junctions of thermocouples $\Delta T$ was fixed
by a recorder with the accuracy class $\sim0.4\%$. The tape advance of
$25cm$ corresponded to $1mV$ of thermoelectrical voltage, i.e. to $\Delta T\approx25^\circ$.
The more thermocouples at point (8) gave a growth in sensitivity of the device
but in the value of random error, too.

The plasma generator current and voltage were read with
an accuracy of $1.5\%$ of the limiting values for the instruments
used ($750A$ and $500V$, respectively). The flow rate of water
through the cooled measuring tube was fixed to within $3\%$ ($\sim 60g/s$).
The mass velocity $V$ of particles in the plasma generator
jet was equal to $120m/s$. The ionization coefficient in the jet
was $\sim0.1\%$.  The initial experiments were performed in the
following manner. First, a point in time was chosen at which the
vector ${\bf A_g}$ was close to the horizontal plane. Further, the
starting direction of plasma generator jet was set up at an
arbitrary angle to the presumed direction of ${\bf A_g}$.  When the
plasma generator was in the operating conditions and the
readings from the thermocouples corresponded to a stationary
regime ($\Delta T = Const$), one began to turn the plasma generator in
the horizontal plane together with the whole instruments around
the vertical axis at $< 5^\circ$ a second. At this instant, the recorder
fixed the value of the angle and the corresponding $\Delta T$.

When investigating, a hypothesis that the indications of the thermocouples
$\Delta T$ were independent on the angle of rotation $\varphi$, was assumed.
The deflections of $\Delta T$ during the experiment from its stationary value
found before rotation, which are not explicable in the context of the traditional
physics were considered as manifestations of the new force action. To analyze a
result, the experimental values of $\Delta T$ were averaged over $10^\circ$-sectors
of rotation, and the stationary value of $\Delta T$ (before rotation) was checked
during $\sim 2 min$ before each experiment, i.e. nearly for an assumed time
of rotation of the plasma generator in the process of the experiment.
The plasma generator used was able to operate in the stationary mode during $\sim 30 min$.

It should be noted that when rotating the plasma generator, the turning radius
of the hose conducting water to the measuring tube was fixed near the latter
and did not exceed 20 cm. The system of all water hoses and the power cables
was untwisted in the process of the plasma generator rotation, i.e. the radii
of their curvature increased and tended to infinity.

\begin{figure}[thb]  
\centerline{\psfig{figure=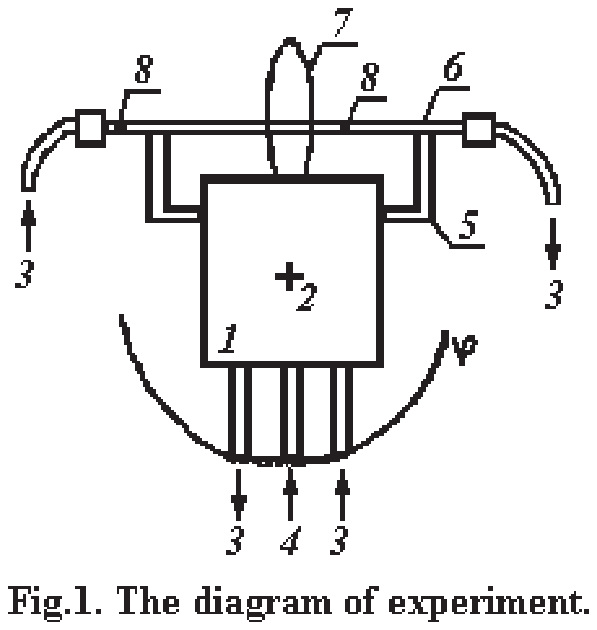}}
\vspace{-5mm}
\end{figure}
\begin{figure}[thb]  
\centerline{\psfig{figure=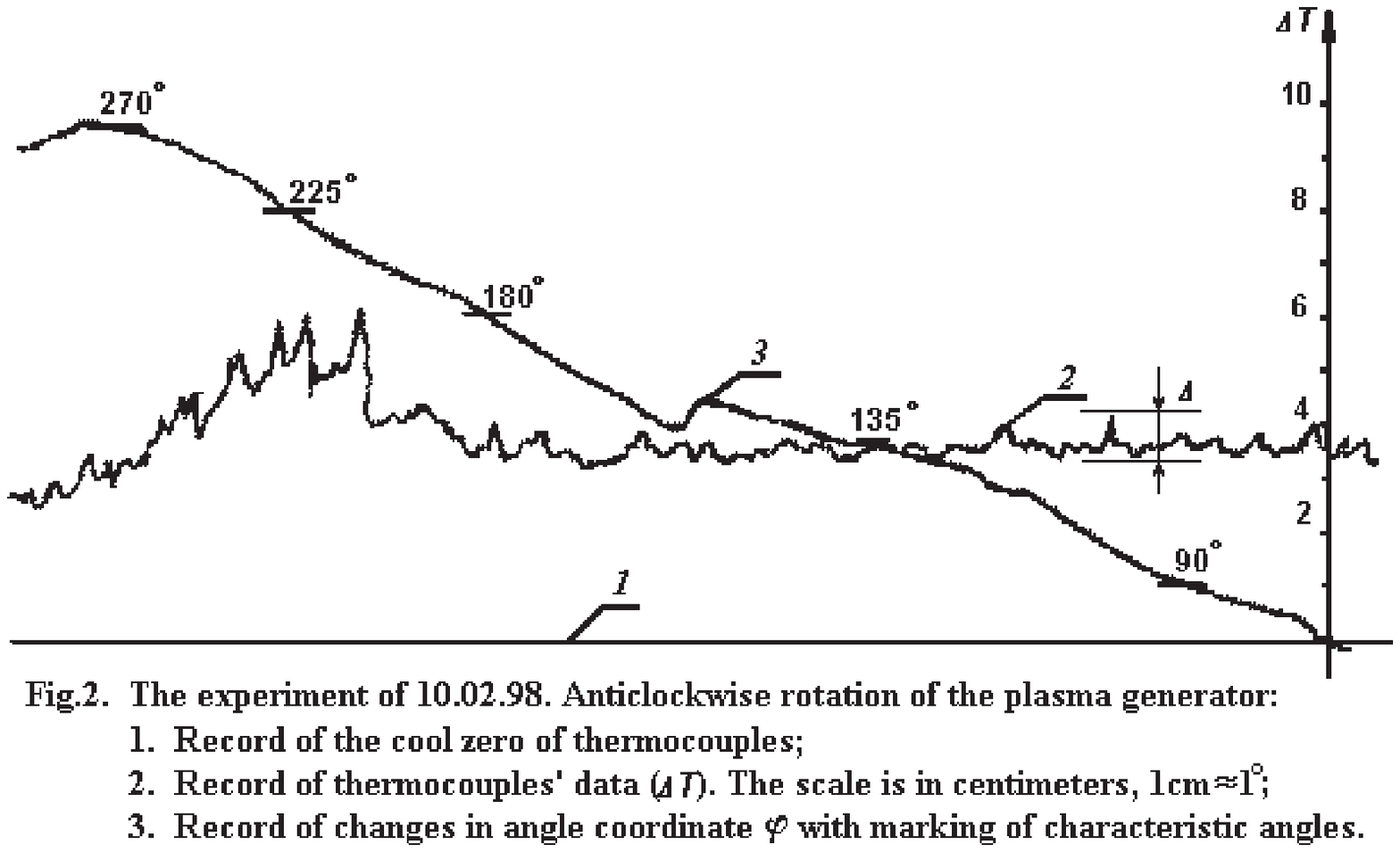,width=135mm}}
\vspace{-5mm}
\end{figure}
\begin{figure}[thb]  
\centerline{\psfig{figure=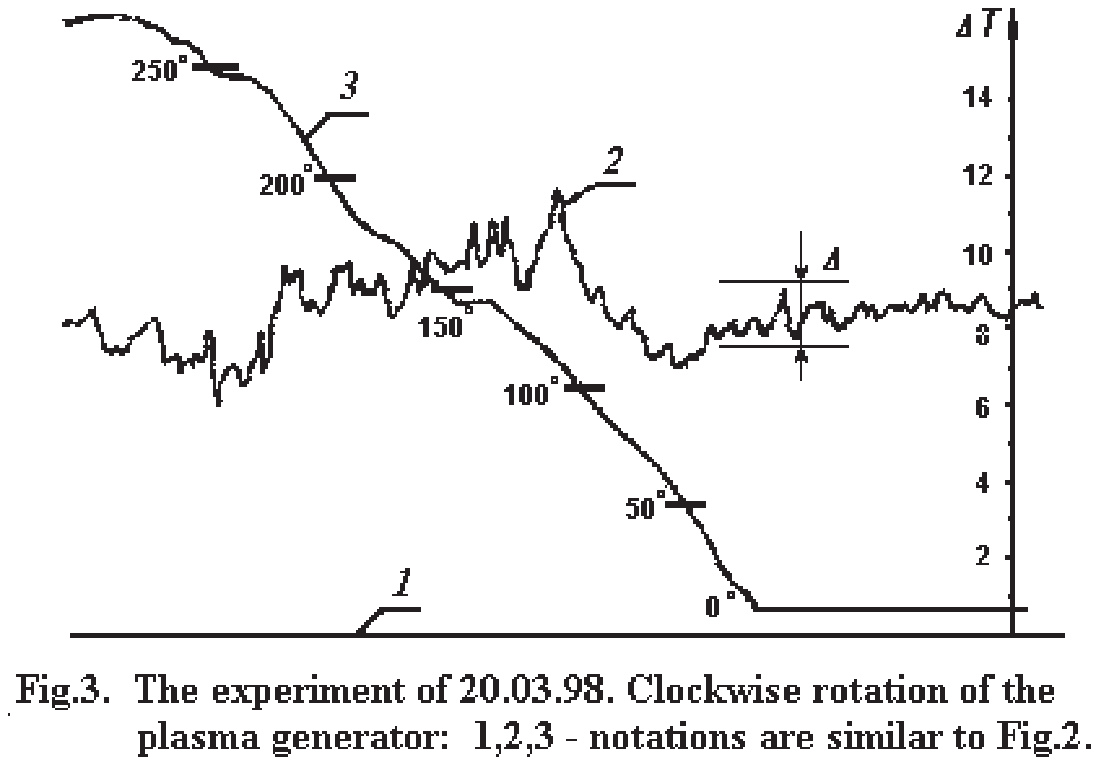}}
\vspace{-5mm}
\end{figure}

\vskip10pt
{\bf 3. The results of experiments.}
\vskip5pt
In Figs. 2 and 3 the results of
two experiments carried out in February 10,1998, at $9^{50}$, and in
March 20, 1998, at $7^{45}$ (Moscow time), are shown. In the first of
them the plasma generator was turned counterclockwise from some
arbitrary angle. As is seen from Fig.2, a considerable increase
in $\Delta T$ reaching $40\%$ of the stationary condition was observed
close to an angle of $225^\circ$. In Fig.3 we also see a substantial
rise in  $\Delta T$ ($\sim 35\%$) nearby $120^\circ\div150^\circ$ when turning the plasma
generator clockwise beginning from another arbitrary angle. In the first
experiment, the turn angle of the plasma generator equal to $225^\circ$
corresponded to the angle $\alpha\sim 340^\circ$ of the maximum action of the
new force. In the second experiment this maximum action
corresponded to the average value $\alpha\sim 260^\circ$. Because the angles
of maximum action of the new force lie on the left and on the
right of the vector ${\bf A_g}$, and the force along the direction of ${\bf A_g}$
itself is zero [8], the initial experiments have given the
direction of ${\bf A_\Sigma}$ with the coordinates $\alpha\sim 300^\circ, \delta\sim 34^\circ$
(the vectors ${\bf A_g}$ and ${\bf A_\Sigma}$ are almost parallel). Therewith the angle
between the vector ${\bf A}$ of the current in the plasma generator
discharge and the vector ${\bf A_g}$  for maximum action of the new force
was $140^\circ$ for the first and $130^\circ$  for the second experiments,
respectively.

In the second run of experiments carried out by day since June
29, till July 02,1998, the optimum angle $\beta$ the jet made to the
horizontal plane, was sought because at that period the vector
${\bf A_g}$ was nearly horizontal only by nights. At $\beta = 0$ and when
turning the plasma generator around the vertical axis through
$320^\circ$, the new force did not manifest itself at all. As the angle
$\beta$ was spaced at $15^\circ$ intervals, at $\beta = 30^\circ$ the maximum inflection
of the $\Delta T$ - curve during rotation of the plasma generator around
the vertical line was observed. In various experiments these
inflections corresponded to the following coordinates of the
maximum action of the new force: $\alpha \approx 255^\circ$,
$\alpha \approx 340^\circ$, and $\beta \approx 30^\circ$ ($\beta\approx\delta$).
That is, the coordinates of the vector
${\bf A_g}$ practically had not changed and were equal to
$\alpha \approx 297^\circ$ and $\beta \approx 30^\circ$ ($\beta\approx\delta$).

Altogether there were carried out more than 20 experiments in 1998. All of
them revealed (with an accuracy of $\sim 20^\circ$) only two directions in
space relative to the vector ${\bf A_g}$ corresponding to maximum $\Delta T$
(see above). The summary statistic error including also random
non-controllable processes (in the discharge of the plasma
generator, in the flow of water nearly the thermocouples in the
measuring tube etc.) was no more then $\pm12\%$ in each individual
experiment on determining $\Delta T$. The latter is clearly seen in
Fig.2.  It is necessary to note that this result was obtained
not only in the experiments with rotation of the plasma
generator from an arbitrary angle but five months later as well
(when the Earth turned through $\sim150^\circ$ about the Sun).  In the
course of the experiments, the bendings of water and gas hoses
were insignificant and did not influence the experimental
result. The action of the Coriolis force was unimportant, too.
It is interesting to note that the results of experiments
carried out in February 1998 and 1999 at the same days and
hours, are qualitatively coincident (with an accuracy of $\sim 20\%$
and with some common turn of the whole field of directions of
the new force through $\sim 20^\circ$). As in 1998, in 1999 also two
directions of this force with a difference in $\alpha$ - coordinate
equal to $\sim 90^\circ$, were prominent. For the experiments of February
1999, the coordinates of the vector ${\bf A_g}$ calculated by the same
procedure are $\alpha \approx 280^\circ$ and $\beta \approx 30^\circ$.
It should be pointed out that
the manifestation of the new force on the left or the right of
${\bf A_g}$ was accidental, i.e. we could not precisely predict when this
force will be fixed by us in the process of turning the plasma
generator: before or after the passage of jet direction through
the presumed direction of ${\bf A_g}$. In roughly 30 experiments
performed by us the force manifested itself after that passage
approximately twice as frequently as before.

It is also interesting to note that in some experiments, an increase in $\Delta T$
at some angles $\varphi$ was immediately followed by one third as many decrease
in its amplitude, - see Fig.1, $\varphi \approx 270^\circ$).

In Fig.4 shown are (in the projection onto the plane of celestial equator,
$\delta = 0$) the direction of action of the new force $F$ and that of the
vector ${\bf A_g}$, determined from $\Delta T$ change in the plasma generator
jet, for typical experiments performed in various day times and
months of the year (the direction of the new force for other experiments are
within the range of its direction shown in Fig. 4). As is seen, the new force directions in
space are obviously not accidental, two of them (indicated above) are prominent.
In the following section a detailed analysis of experimental errors is given.

\begin{figure}[p]  
\centerline{\psfig{figure=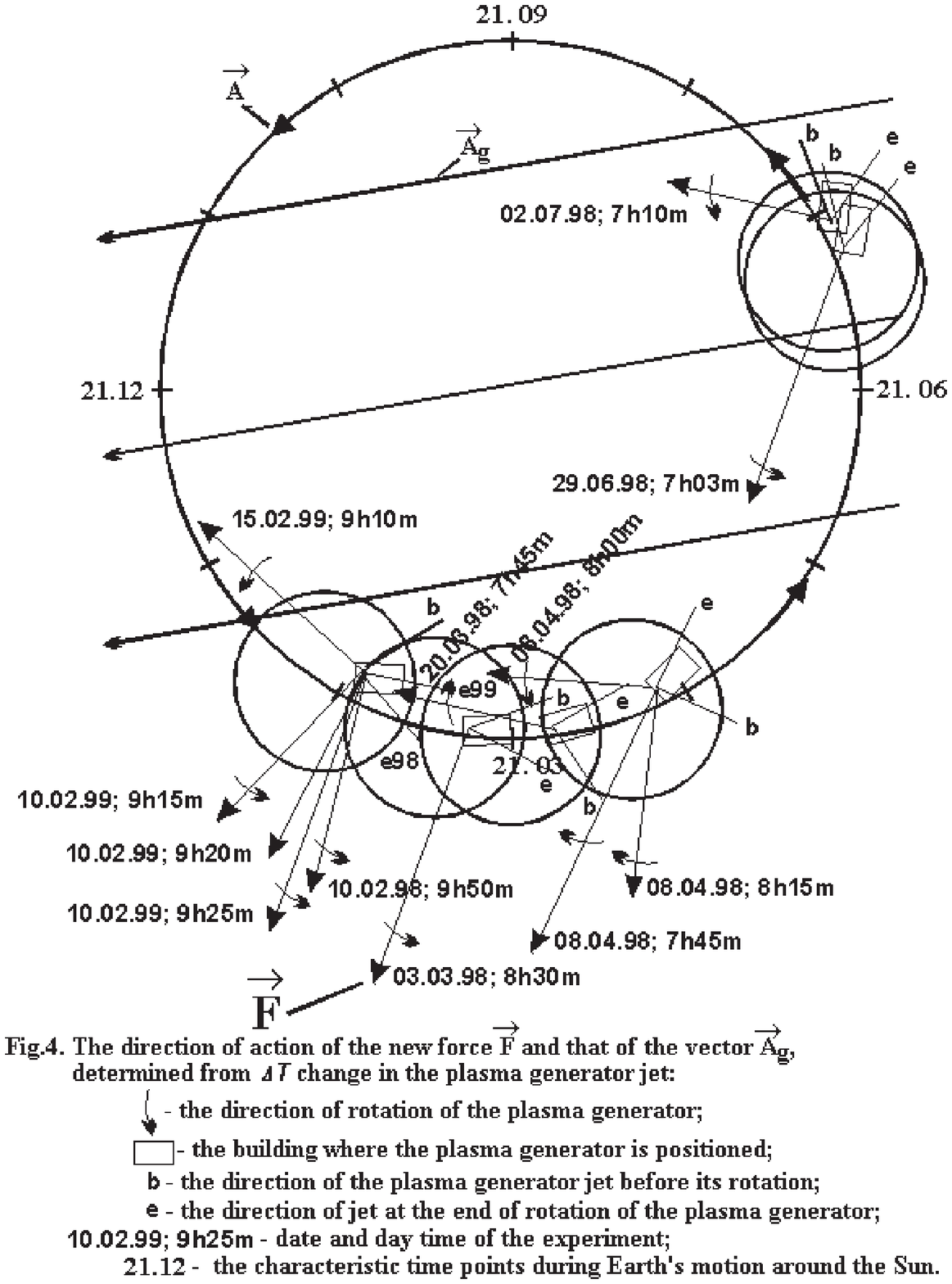,angle=0,height=190mm}}
\vspace{5mm}
\end{figure}

{\bf 4. Analysis of errors in the experiment}
\vskip5pt
The errors in determining the extremum of $\Delta T$ versus $\varphi$ are classed as
systematic $\sigma_{syst}$ and random (statistic) $\sigma_{stat}$ ones.
The systematic errors could be caused by the curvature of hoses,
wrong position of thermocouples relative to the jet, Coriolis force,
zero drift in readings of thermocouples, their calibration.

The statistic error $\sigma_{stat}$ was caused by the random processes in the discharge of the plasma
generator in the measuring tube (when flowing around thermocouple junctions by water filaments
with different temperature), finite accuracy of measuring devices (ammeter ($\sigma^A_{stat}$), voltmeter ($\sigma^V_{stat}$),
flowmeters of water in the measuring tube ($\sigma^{W_1}_{stat}$) and in the plasma generator cooling system
($\sigma^{W_2}_{stat}$), flowmeters of air and argon ($\sigma^{Ar}_{stat}$)) and of recorders ($\sigma^{R}_{stat}$) as well as by the
experimental conditions (background): uncontrollable convective processes (wind),
electromagnetic situation in the room.

Consider the systematic errors.

Preliminary investigations on the importance of bend radii of the hoses conducting water to
the measuring tube and for cooling the plasma generator, have shown the following. If the radii of
the water hoses close to the measuring tube were fixed at the value of 20 cm, and the remainder
of the "measuring" and "cooling" hoses were untwisted in the process of rotation so that their
bend radii tended to $\infty$, then the change in the pressure loss owing to bending of the hoses did not
tell on the value of the summary error in determining $\Delta T$ during the rotation with an accuracy
better that $-3\% (\sigma^H_{syst})$.

A dramatic effect on the valid signal to noise ratio was made by the position of thermocouple
junctions relative to the jet. The maximum signal was obtained when one thermocouple was in the
vicinity of the jet and the other was at the inlet of the measuring tube. When the thermocouples
were placed symmetrically relative to the jet (at a distance of 30 cm between them), the effect
became considerably weaker.

The above said is easily seen from Table 1 where the summary error and the valid signal are
given. In the experiments 11 and 12, an old measuring tube was used. To prevent it from
burn-out, we removed the thermocouple from the jet zone for a distance of 5 cm. The valid signal was
weaker in that case. The effect in question was studied and reduced to some corrections to
experimental results.

The Coriolis force also had a great effect on the deflection of $\Delta T$ from the stationary value.
We observed an increase in $\Delta T$ owing to this force during anticlockwise rotation and a decrease in
$\Delta T$ during clockwise rotation  (see Fig.3, $\varphi\approx 50^\circ$).
The investigation of this influence led to a limitation on the plasma generator
rotation velocity around its axis. A value of the maximum angular velocity was
found ($\sim 5$ degrees per second) at which the deflections of $\Delta T$-curve owing
to the Coriolis force at clockwise and anticlockwise rotation were no more than $3\%$
(if the flow rate kept constant). The action of the Coriolis force was
thoroughly studied, too, and gave the above-mentioned correction to the final result.

In some experiments (Table 1, experiments 11-16), an insignificant ($\sim 7\%$) pure linear
zero drift in the readings of thermocouples was observed. The drift was caused by heating of
water in the closed tank from which the intake for the measuring system was made. The said drift
was also taken into account as a correction in the analysis of the final result.
Thus $\sigma_{syst}\approx\sigma^H_{syst}\approx-3\%$.


Let us estimate random error.

For the above indicated parameters of the discharge, the random errors were equal to $\pm3,5\%$
in current ($\sigma^A_{stat}$) and $\pm3,2\%$ in voltage $\sigma^V_{stat}$.
As they were not independent, the summary error of the power system were
$\sigma^W_{stat} = \pm 6,7\%$. The further random errors were those of calibration of
thermocouples ($\sigma^T_{stat} \approx \pm 0,3\%$), of the water flowmeter
in the measuring tube ($\sigma^{W_1}_{stat} \approx \pm 1\%$) and in
the cooling system of the plasma generator ($\sigma^{W_2}_{stat} \approx \pm 3\%$),
of the flowmeters of air and argon in common ($\sigma^{Ar}_{stat} \approx \pm 4\%$),
of the recorder ($\sigma^{R}_{stat} \approx \pm 0,4\%$).

Consider experimental conditions. The experiments were carried out in a great room $30\times40 m$
in area and $15 m$ in height with controllable temperature, which remained constant in the course of
experiments in spite of operation of the plasma generator. In the vicinity of the latter there were
no objects acting on the character of plasma jet. The room was free from a noticeable wind. The
convective flows initiated by the plasma generator were not monitored. In the neighboring rooms
as well as in the experimental room one did no works with electromagnetic devices which could
influence on the results of experiments with the plasma generator. The possible effect of magnetic
storms changing the magnetic field in the room by $\sim 10^{-3} Oe$ could be neglected for the experiments
in consideration. Based on data of IZMIRAN institute for the Earth's magnetic surrounding,
it were not marked any magnetic storms per the days the investigation was carried out, specified in Table 1.

In connection with the independence of the above-listed factors giving random errors, the
total mean-root-square error was equal to
$$\sigma^{\Sigma}_{stat} = \sqrt{(\sigma^W_{stat})^2 + (\sigma^T_{stat})^2 +
(\sigma^{W_1}_{stat})^2 + (\sigma^{W_2}_{stat})^2 + (\sigma^{Ar}_{stat})^2 +
(\sigma^{R}_{stat})^2} \approx \pm 8\%.$$
In the course of experiments we had the summary error including all errors, as systematic so
random ones, except only $\sigma^{R}_{stat}$ which was very small.
The analysis of data in Table 1 and the value of $\sigma^{\Sigma}_{stat})$ shows
that in some experiments $\sigma^{\Sigma}_{stat} > \sigma_{\Sigma}$,
which can be probably explicated by compensation of errors when determining $\sigma^{\Sigma}_{stat}$.

The errors connected with determining the vector ${\bf A_g}$ and caused by averaging the extremum
values of the curve $\Delta T(\varphi)$ and by construction of vectors of the new force, added up to no more
than $\pm 10\%$.
\vskip5pt
\noindent
{\bf Table 1.}  The values of the summary error $\sigma_{\Sigma}$ in 20 experiments, 1998-99.
\begin{center}
{\small
\noindent
\begin{tabular}{|c|c|c|c|c|}
\hline
N &Date and time                  &Angle        &Summary      & Deflection of $\Delta T$   \\
&of the experiment&$\beta^\circ$&error $\sigma_\Sigma$&by the new force $F$ \\
\hline
1.&       10.02.1998 9h 50min  &         0  &  $\pm 12\%$ &    40\% \\
\hline
2.& 	  3.03.1998 8h 30min &          0   & $\pm3\%$   &  10\%\\
\hline
3.& 	  3.03.1998 9h 00min &          0 &   $\pm3,5\%$   &      13\% \\
\hline
4.& 	  3.03.1998 9h 30min &          0 &   $\pm5\%  $   &     15\%  \\
\hline
5.& 	  20.03.1998 7h 45min&          0 &   $\pm9\%  $   &     35\%  \\
\hline
6.& 	  20.03.1998 8h 45min&          0 &   $\pm10\% $   &     30\%  \\
\hline
7.& 	  24.03.1998 8h 30min&          0 &   $\pm5\%  $   &     21\%  \\
\hline
8.& 	  24.03.1998 9h 15min&          0 &   $\pm5\%  $   &     18\%  \\
\hline
9.& 	  4.04.1998 7h 45min &          0 &   $\pm5,5\%$   &     14\%  \\
\hline
10. &	  8.04.1998 8h 15min &          0 &   $\pm5,5\%$    &     12\%  \\
\hline
11. &	  22.06.1998 7h 10min&          15&   $\pm5\%  $   &     10\%  \\
\hline
12. &	  22.06.1998 7h 50min&          15&   $\pm5\%  $   &     8\%   \\
\hline
13. &	  29.06.1998 7h 03min&          30&   $\pm6,5\%$   &     17\%  \\
\hline
14. &	  29.06.1998 7h 45min&          30&   $\pm5\%  $   &     12\%  \\
\hline
15. &	  2.07.1998 7h 10min &          30&   $\pm11\% $   &     27\%  \\
\hline
16. &	  2.07.1998 8h 15min &          30&   $\pm10\% $   &     28\%  \\
\hline
17. &	  10.02.1999 9h 15min&          0 &   $\pm3\%  $   &     10\%  \\
\hline
18. &	  15.02.1999 9h 50min &          0 &  $\pm 6\% $    &     14\%  \\
\hline
19. &	  16.02.1999 9h 10min&          0 &   $\pm2,5\%$   &     8\%   \\
\hline
20. &	  16.02.1999 10h 15min&          0 &  $\pm 2,5\%$    &    10\%   \\
\hline
\end{tabular}
}
\end{center}
\pagebreak
{\bf 5. Addendum}
\vskip10pt

In the Central Research Institute of Machinery, the
thermophysical properties of constructional materials in the jet
of a plasma generator with linear discharge $1MW$ in power, were investigated for more
than ten years. In some instances, while estimating the heat
content of the jet with the aid of a local calorimeter, a
considerable excess of energy released in the jet above the
energy taken from the power source, was observed at fixed
operating parameters of the plasma generator ($I = (500\div1800) A; V =
(3000\div3500) m/s$). The results of estimation of the total energy at
the output of the plasma generator were obtained by way of
computations.

In three last experiments (two of them were
carried out in Oct. 22, 1992, at $14^{30},15^{00}$, and one was in Apr.
22, 1994, at $15^{00}$), integral calorimetric measurements in the
jet of the $1MW$ plasma generator were fulfilled by means of a
non-stationary calorimeter crossing the jet in a matter of 0.2
second. The duration of stationary operation of the plasma
generator was equal to $30-40s$, the error of measurements was $\sim \pm20\%$.
The ratio of energy output $W_1$ to input $W_0$ equaled $\sim 1$
in the experiments of 1992 but the measurements in 1994 have given
$W_1/W_0 \approx 2$ which was much more than the error of the experimental
technique used. The latter fact also lent an impetus to
conducting the above described experiments with plasma
generator. An analysis of spatial arrangement of the axis of
plasma generator relative to the vector ${\bf A_g}$ has shown that in the
experiments of 1992, the effect of increasing energy in the jet
was to be totally absent but in the experiment carried out in
1994, by contrast, the plasma generator jet was just at the most
efficient angle ($\sim 30^\circ$) to ${\bf A_g}$. Therewith the angle between the
vector A of the current of the plasma generator and the vector
${\bf A_g}$ was equal to $\sim 150^\circ$.

Since 1976 till 1982 in the Research Center of High-Voltage Equipment (Moscow), a run of experiments
was performed by V.P.Ignatko and others on investigation of
alternating high current electric arc in the closed volume of
transformer oil in a unique experimental set-up having no
analogues in the world. The root-mean-square current was varied
through a range of $20-130kA$ with the amplitude values no more
than $200kA$. The period of current oscillations was equal to
$0.02s$, the duration of arc discharge was $\sim 1s$. The electric
set-up was made up of two shock-exited electric machine
oscillators of {\bf TI-100-2} type in a double transformation circuit,
and delivered up to $12kV$ of r.m.s. no-load voltage. The oil of
volume $(0.6\div1)m^3$ was inclosed in a vertical cylindrical tank from
steel $\sim3m$ in height and $\sim1m$ in diameter weighing about $7t$.
The arc burned in oil between hemispheric copper electrodes $7cm$ in
diameter initially placed $(0.5\div5)cm$ apart in the middle plane of
the cylinder between its bottom and cover.  The current,
voltage, power and energy of the arc, pressure at various points
of the experimental model, and deformations of the whole
construction investigated, were measured. The process was
filmed.  More than 100 experiments were carried out. In some of
them ($\sim8$) anomalous phenomena were observed - a tendency to current
suppression without apparent reasons for that.  So, in an
experiment in Oct. 7, 1976, at $22^{00}$, the current at its third
and fifth half- periods decreased from $186kA$ to $8.7kA$ and $3.3kA$
(i.e. tens times) until the arc decayed 0.05 second after. In a
time of $0.05s$, $5.55MJ$ of energy were transferred to the arc.
That experiment was finished by an accident. Uncontrollable
energy release in the arc took place, and the pressure in the
vertical direction was built up to $120atm$ which terminated in
deflection of the cover of the cylinder ($8cm$ in thickness) by
1cm and emergence of a crack of width up to $0.2cm$. An analysis
of the process have shown that the energy released in the arc
turned out to be about an order higher than the expenditure  of
energy.  The experiment performed in June 10, 1982, at $20^{35}$, in
which the arc current equaled $38.5kA$, the energy imparted
accounted for $31.2MJ$, the initial gap between the electrodes was
equal to $\sim2cm$ (these are common, far from limiting parameters
for the set up in question), led to an explosion. The
seven-tonne cylinder broke away from screw anchors and rose up
having destroyed the ceiling. Examination of deformations
occurred and estimation, on their basis, of arc energy by a
joint commission have shown that there were released in the arc
10-100 times more energy than was communicated to it. The
government commission could not find reasons for the accident in
the limits of existing physical and chemical knowledge.
Analysis of spatial arrangement of the center line of the set-up
electrodes has led to a conclusion that in the latter case it
made the most efficient angle with the vector ${\bf A_g}$ (for the action
of the new force), and in the experiments of 07.10.76 the angle
between that line and ${\bf A_g}$ was very close to the most efficient.
The above mentioned phenomenon of abnormal suppression of
current, inexplicable on the basis of the existing electrical
engineering and physics, also can be explained by the action of
the new force.

The last of 1999, we have been acquainted with the works of R.M.Santilli
\cite{22} in which, on the basis of a new model of unstable hadrons developed
by him ("the hadronic mechanics"), a Santilli's reactor was proposed using
an arc discharge in a special liquid. The power output in experiments
with the reactor turned to be much more than the energy taken from the
power source.

In the present paper we shall not consider distinctions between two
physical models ("the hadronic mechanics" and the new interaction connected
with the existence of ${\bf A_g}$). We note only that the arc discharge in
a special liquid can lead to formation of vector-potential lines of
thoroidal type, in which case the new interaction will manifest
itself in no dependence upon the day time since in the thoroid there are
always the lines of the vectorial potential directed towards the vector ${\bf A_g}$.

Therefore our experiments and those of R.M.Santilli  have possibly the same
physical base.

Thus the material of the present paper and the
whole complex of investigations of properties and
characteristics of the new force as well as of the global
anisotropy of space due to existence of the vector ${\bf A_g}$, testify
that we have detected a new source of energy connected to the
energy of the physical space. This energy can be used with the
aid of various current-carrying systems acting by potentials on
elementary particles through which the energy comes to us.

\vskip10pt
{\bf Acknowledgments}
\vskip10pt

The authors are sincerely grateful to the academicians
A.M.Prokhorov, S.T.Belyaev, V.F.Utkin, N.A.Anfimov,
G.E.Losino-Losinsky for support of the work and useful
discussions, as well as to A.A.Rukhadze, I.B.Timofeev,
V.B.Fyodorov, Yu.L.Sokolov and many other participants of
scientific seminars at the Institute of General Physics of RAS, Moscow State
University and Scientific Center "Kurchatovsky Institute" for
fruitful discussions and useful advices.

\pagebreak

\end{document}